\documentclass[num-refs]{wiley-article}

\usepackage{siunitx}

\papertype{Original Article}
\paperfield{Journal Section}

\title{An Ultra-Sub-Wavelength Microwave Polarization Switching Antenna for Covert Communication Implemented with Surface Acoustic Waves in an Artificial Multiferroic Magnonic Crystal}

\author[1]{Raisa Fabiha}
\author[1]{Erdem Topsakal}
\author[1]{Supriyo Bandyopadhyay}

\affil[1]{Department of Electrical and Computer Engineering, Virginia Commonwealth University, Richmond, VA 23284, USA}

\corraddress{Supriyo Bandyopadhyay, Department of Electrical and Computer Engineering, Virginia Commonwealth University, Richmond, VA 23284, USA}
\corremail{sbandy@vcu.edu}

\fundinginfo{Virginia Commonwealth University Commercialization Fund, Commonwealth Cyber Initiative Faculty Fellows Program}

\runningauthor{Raisa Fabiha et al.}

\begin{document}

\begin{frontmatter}
\maketitle

\begin{abstract}
The ability to switch at will the polarization of a transmitted electromagnetic
wave from vertical to horizontal, or vice versa,
is of great technological interest because of its many applications
in long distance communication (e.g., polarization division multiplexing). Binary bits can be
encoded in the two orthogonal polarizations and transmitted from point to point. Polarization switches, however,
are usually much larger than the wavelength of the
electromagnetic wave that they transmit. Consequently, most research in this
area has focused on the optical regime where the wavelength
is relatively short ($\sim$1 $\mu$m), so that the switch being
much larger than the wavelength is not too inconvenient.
However, this changes in the microwave regime where the
wavelength is much larger (typically $>$ 1 cm). That makes a
microwave {\it ultra-sub-wavelength} polarization switch very attractive.
Here, we report, for the first time to our knowledge, such a switch made of an array
of magnetostrictive nanomagnets ($\sim$100 nm lateral dimension, $\sim$5 nm thickness)
deposited on a piezoelectric substrate to make an ``artificial multiferroic
magnonic crystal (AMMC)''. A surface acoustic wave (SAW)
launched in the substrate with suitable electrodes excites confined
spin waves in the nanomagnets via phonon-magnon coupling, which then
radiate electromagnetic waves in space via
magnon-photon coupling. In some particular direction(s), determined by the AMMC parameters, the polarization of the beam at a given frequency can
be rotated through $\sim$90$^{\circ}$ by switching the direction of SAW
propagation in the piezoelectric substrate between two mutually
orthogonal directions via activation of two different pairs of launching electrodes. By aligning the transmitter and
the receiver along such a  direction (known only to
authorized users), one can communicate {\it covertly} from point
to point, without the need for encryption or cryptography. Furthermore, this attribute also makes the antenna ``stealthy'' since the message can be concealed from any eavesdropper whose receiver is not precisely aligned in the correct direction.
\end{abstract}
\end{frontmatter}

\section{Introduction}
In quantum communication, information is encoded in the polarization of a photon for quantum key distribution and other tasks that call for an unconditionally secure link. In classical communication, encoding information in the polarization of an electromagnetic wave offers some advantages as well, such as polarization encoded secret sharing \cite{oe,li} and  multiple data streams sent on the same frequency channel using different polarization states to save bandwidth. The latter is known as {\it polarization division multiplexing} which is used in satellite communication but could be also used in on-chip communication to reduce both bandwidth and component count.

For {\it digital} data transmission via two orthogonal polarizations encoding bits 0 and 1, one would require a polarization switch that will switch the polarization of an electromagnetic wave from approximately horizontal (encoding the bit 0) to approximately vertical (encoding the bit 1), or vice versa. Absolute polarization purity is not a concern for such applications. As long as the two polarizations are distinguishable from each other (i.e., they are approximately orthogonal), it will suffice. 

Microwave polarization switches are however rare since they are usually much larger than the microwave wavelength (1-10 cm) and hence not integrable on a chip, although there are some exceptions \cite{staacke,jin}. Consequently, most  polarization switches work at optical frequencies where they are still larger than the optical wavelength ($\sim$1 $\mu$m) but small enough to be integrated on a chip. Optical polarization switches are implemented with various techniques such as rotating waveplates \cite{imai}, Babinet-Soleil compensators \cite{collett}, Berek rotary compensators \cite{holmes}, fiber coil polarization controllers \cite{lefevre}, Faraday rotators \cite{okoshi},  degree of polarization generators \cite{lizana}, lithium niobate electro-optics \cite{kubota}, liquid crystals \cite{zhuang}, digital micromirrors \cite{she}, graphene metasurfaces \cite{farzin}
and on-chip photonic circuits \cite{rodrigues,dong,miller}. All of them are typically much larger than the wavelength, but that is hardly a problem in the domain of optics where the wavelength is around only 1 $\mu$m. It becomes a serious problem in the microwave frequency region where the wavelength can be several cm. Therefore, sub-wavelength polarization switches are very attractive for the microwave regime. Here, we demonstrate a microwave polarization switch antenna (working at 1-6 GHz) that  is orders of magnitude {\it smaller} than the wavelength. This allows integration on a chip. That can have multiple applications, such as in quantum computing \cite{staacke,brown}, remote sensing \cite{tyo} and, most importantly, aggressively miniaturized  ultra-compact polarization division multiplexers and digital data transmitters in the microwave range. Additionally, it turns out that it may have other important applications, such as in secure communication in embedded applications (e.g., medically implanted devices, tiny stealthy drones for defense and crime-fighting), stealthy antennas to operate in hostile environments and physically unclonable functions (PUF) for antenna authetication and trust.

 \begin{figure}[hbt!]
\centering
\includegraphics[width=0.99\textwidth]{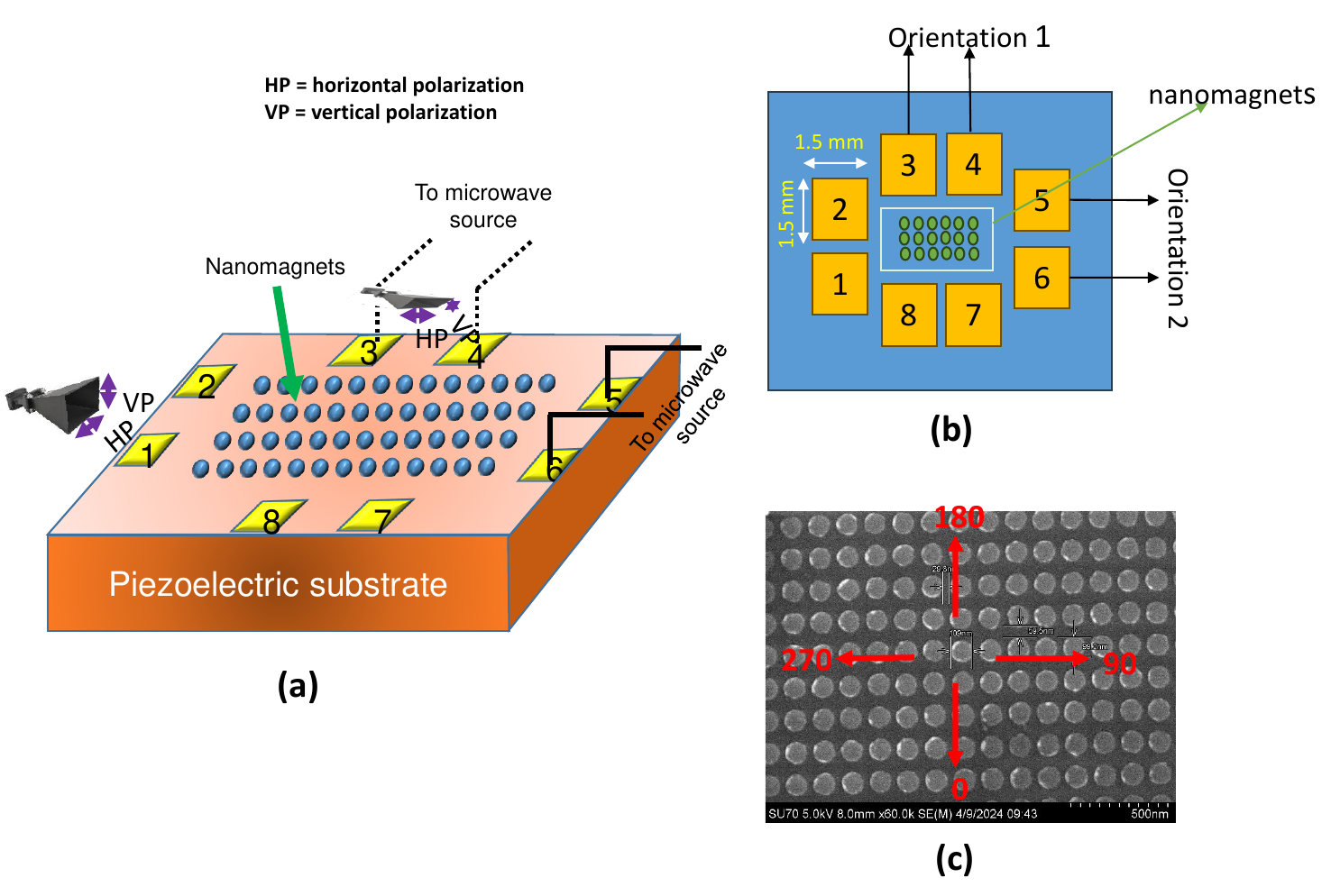}
\caption{(a) Schematic of the ultra-sub-wavelength microwave polarization switch. A two-dimensional periodic array of magnetostrictive nanomagnets is delineated on a piezoelectric substrate. The different pairs of electrodes, such as (3,4) or (5,6) can be connected to a microwave voltage source to launch surface acoustic waves (SAWs) in different directions within the substrate. This radiates electromagnetic waves in space. The polarization of the radiated electromagnetic beam in any given direction at any given microwave frequency depends on the direction of SAW propagation and can be changed by changing the latter via switching between different pairs of electrodes. When the radiation pattern is measured in the plane of the nanomagnets, ``horizontal'' polarization refers to in-plane polarization and ``vertical'' polarization refers to out-of-plane polarization. (b) Top view of the device. Here, ``orientation 1'' is defined as the case when the microwave
source is connected between electrodes 3 and 4 to launch a SAW propagating parallel to the major axes of the
elliptical nanomagnets, while ``orientation 2'' refers to the case when the microwave source is connected between
electrodes 5 and 6 to launch a SAW propagating parallel to the minor axes of the elliptical nanomagnets. (c)  Definition of the ``directions'' for the radiation pattern; 0$^{\circ}$ is along the minor axes and 90$^{\circ}$ is along the major axes of
the nanomagnets. Reproduced partly from \cite{raisa} with the permission of IEEE.}
\label{fig:layout}
\end{figure}


\section{Results and Discussion}

The structure of the polarization switch is shown in Fig. \ref{fig:layout}(a). It consists of a periodic two-dimensional array of magnetostrictive nanomagnets (made of cobalt) deposited on a piezoelectric substrate (128$^{\circ}$ Y-cut  LiNbO$_3$). The presence of both a magnetostrictive and a piezoelectric component effectively forms a two-phase multiferroic. This is  called an ``artificial multiferroic magnonic crystal'' (AMMC) because of its periodicity. It is the same system that was used in ref. \cite{raisa1} for beam steering. Here, it is used for a different purpose, namely polarization switching. All antenna properties such as the scattering parameter $S_{11}$ spectrum, radiation efficiencies at different frequencies, etc. can be found in refs. \cite{raisa,raisa1} and some are also repeated in the Supporting Information.

The 128$^{\circ}$ Y-cut LiNbO$_3$ substrate is 0.5 mm thick. The cobalt nanomagnets are  slightly elliptical with major axis $\sim$110 nm and minor axis  $\sim$100 nm. The thickness is $\sim$ 6 nm. Each nanomagnet has a 5 nm thick Ti layer underneath for adhesion to the substrate. The nanomagnet array covers an area of $\sim$100 $\mu$m $\times$ 100 $\mu$m and is fabricated with electron-beam-lithography using a Raith Voyager e-beam writer. The substrate is spin-coated with a single layer of PMMA resist spun at 2500 rpm and baked at 110$^{\circ}$C for 2 minutes. It is then patterned with e-beam, and developed in a methyl isobutyl ketone and isopropyl alcohol (MIBK-IPA, 1:3) solution for 60 seconds, followed by a cold IPA rinse. After the patterning is complete, a 5 nm-thick titanium (Ti) adhesion layer is deposited using electron beam evaporation at a base pressure of 2.3 $\times$ 10$^{-7}$ Torr, followed by the deposition of 6 nm thick cobalt. Lift-off is performed using remover PG solution. A scanning electron micrograph of the nanomagnets  can be found in ref. \cite{raisa}. The electrodes for launching surface acoustic waves are square patches delineated with optical lithography. They are made of Al which is $\sim$1 $\mu$m thick and have lateral dimensions of 1.5 mm $\times$ 1.5 mm. 

A surface acoustic wave (SAW) can be launched in two mutually orthogonal directions by activating electrode pairs (3,4) or (5,6) (see Fig. \ref{fig:layout}). The electrode pairs are square patches instead of interdigitated transducers (IDTs) since IDTs are also narrow band filters that do not allow launching of multiple SAW frequencies, which is required. In our case, the SAWs are not the usual Rayleigh waves, but a mixture of Rayleigh, Sezawa, Lamb modes. This does not matter since the only purpose of the SAW is to produce time varying strain in the magnetostrictive nanomagnets and make their magnetizations precess (via the Villari effect) to produce spin waves. The spin waves are confined (quantized) standing waves in the nanomagnets whose power and phase profiles were reported in \cite{raisa,nanoscale}. They radiate electromagnetic waves in space via magnon-photon coupling \cite{raisa,raisa1,saibal}. Here magnon-photon coupling implies that the spin waves transfer their energy to electromagnetic waves via mode coupling and this results in the emission of electromagnetic waves in space (antenna functionality). Unlike a conventional electromagnetic antenna that radiates electromagnetic waves owing to fluctuating charges or time-varying electric dipoles, these structures (nano-antennas) radiate electromagnetic waves due to fluctuating magnetization associated with spin waves \cite{raisa,raisa1,saibal,carman,arxiv1,arxiv2}. Hence, they do not behave as traditional antennas and are not constrained by the Harrington limit \cite{Harrington,skrivervik} or Chu's limit \cite{Chu, UCLA} that afflict the gain and bandwidth of traditional antennas.

In the past, we measured the radiation pattern of these samples by propagating a SAW in two different directions in the piezoelectric substrate \cite{raisa}.  We can apply a microwave frequency voltage between either electrode pairs (3,4) or  (5,6) to launch SAWs in two mutually perpendicular directions. We then measure the radiation patterns in an anechoic chamber for these two different directions of SAW propagation. The patterns are measured in the plane of the nanomagnets and in the two planes transverse to the plane of the nanomagnets. They are measured in an AMS-8701
Anechoic Chamber, Antenna Measurement System using a
3164-10 Open Boundary Quad-ridged Horn Antenna. The sample (antenna) is always placed at a distance of 284.5 cm from the horn antenna which ensures that we are measuring the far-field radiation pattern at all frequencies.

These   radiation patterns can be found in ref. \cite{raisa} and also the Supporting Information. 
They show significant anisotropy (radiation intensity is direction-dependent) at all frequencies measured (despite the antenna  being an ultra-sub-wavelength structure that should behave like a point source which radiates isotropically), but more importantly, the radiation pattern depends on the direction of SAW propagation (the difference between ``orientation 1'' and ``orientation 2''). The difference is, of course, more pronounced at some frequencies than at others. This latter feature is very likely caused by the fact that the strengths of magnon-phonon and magnon-photon couplings are frequency-dependent. A rigorous theory that can explain the frequency dependence is outside the scope of this work and will be explored in the future.

 We carried out micromagnetic simulations with OOMMF and MuMax3 to shed light on the directional dependence of the radiated intensity. These simulations (based on Landau-Lifshitz-Gilbert equations) yield the time evolution of the magnetization in the nanomagnets in the presence of periodic strain due to the SAW, and hence reveal the spin waves that are excited by the SAW. We found that the spin wave patterns are strongly influenced by whether the surface acoustic wave propagates parallel to the major axis of the elliptical nanomagnets (as would be the case if electrodes 3 and 4 in Fig. \ref{fig:layout} were excited) or parallel to the minor axis of the nanomagnets (as would be the case if electrodes 5 and 6 were excited instead; see supplementary material of \cite{raisa1}). Since it is the spin waves that radiate the electromagnetic waves via magnon-photon coupling, the electromagnetic radiation pattern also depends strongly on the direction of SAW propagation \cite{raisa}.

 \begin{figure*}[!hbt]
\hspace{-1in}
\includegraphics[width=0.99\textwidth,angle=270]{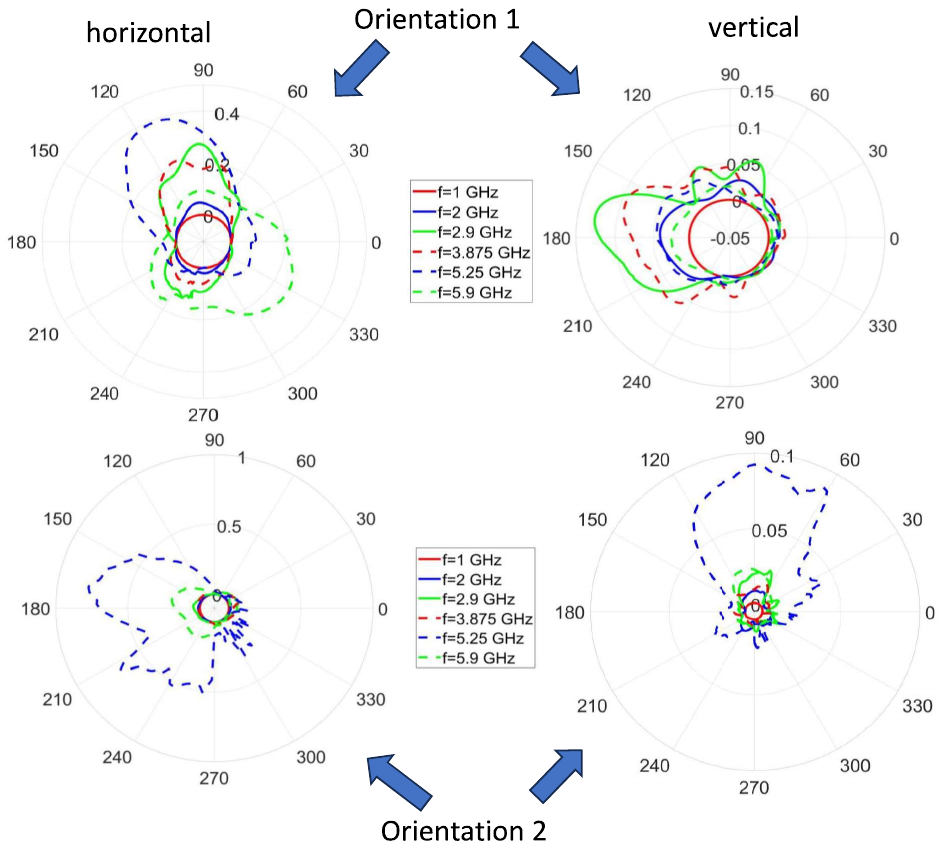}
\vspace{-0.2in}
\caption{Radiation patterns for horizontal and vertical polarizations (gain in absolute units, as opposed to dbi) in the {\it plane of the nanomagnets}  at different SAW excitation frequencies when the microwave source to launch the SAW is connected between two different electrode pairs in order to make the SAW propagate along two different directions designated as ``orientation 1'' and ``orientation 2''.}
\label{fig:Fig3}
\vspace{-0.2in}
\end{figure*}

Why the spin wave pattern depends on the direction of SAW propagation is easily understood. The surface acoustic wave subjects the nanomagnets to periodic strain. Strain acts like an effective magnetic field in a magnetostrictive nanomagnet and in the case of SAW, an effective periodic magnetic field will appear in the nanomagnet \cite{nanoscale}. This periodic field will make the magnetization process and oscillate in time, thereby producing a spin wave. The direction of the effective magnetic field is along the direction of SAW propagation \cite{nanoscale}. Therefore, if we change the direction of SAW propagation, we will change the direction of the effective time-varying magnetic field in the nanomagnet and this will change the axis of precession of the magnetization within the nanomagnet, resulting in a change in the oscillations of the x-, y- and z-components of the magnetization, i.e. the spin wave pattern will change \cite{raisa1}. This then will change the emission intensity of electromagnetic waves in different directions, resulting in a change in the radiation pattern. Ref. \cite{raisa} reported the dependence of the radiation pattern on the direction of SAW propagation for both horizontal and vertical polarizations, while ref.\cite{raisa1} found that changing the direction of SAW propagation changes the radiation {\it spectrum} as well. The latter is due to the fact that when the spin wave patterns change, their frequency components change too and that changes the spectrum of the emitted electromagnetic radiation.

  Changing the spin wave patterns in space by changing the direction of SAW propagation also {\it changes the polarization of the radiated beam in any direction}. This was not studied earlier and here we report that study. A phenomenolgical theory to explain why the polarization will depend on the direction of SAW propagation is provided in the Supporting Information.

\begin{figure*}[!hbt]
\vspace{-0.2in}
\hspace{-0.1in}
\centering
\includegraphics[height=6.0in,angle=270]{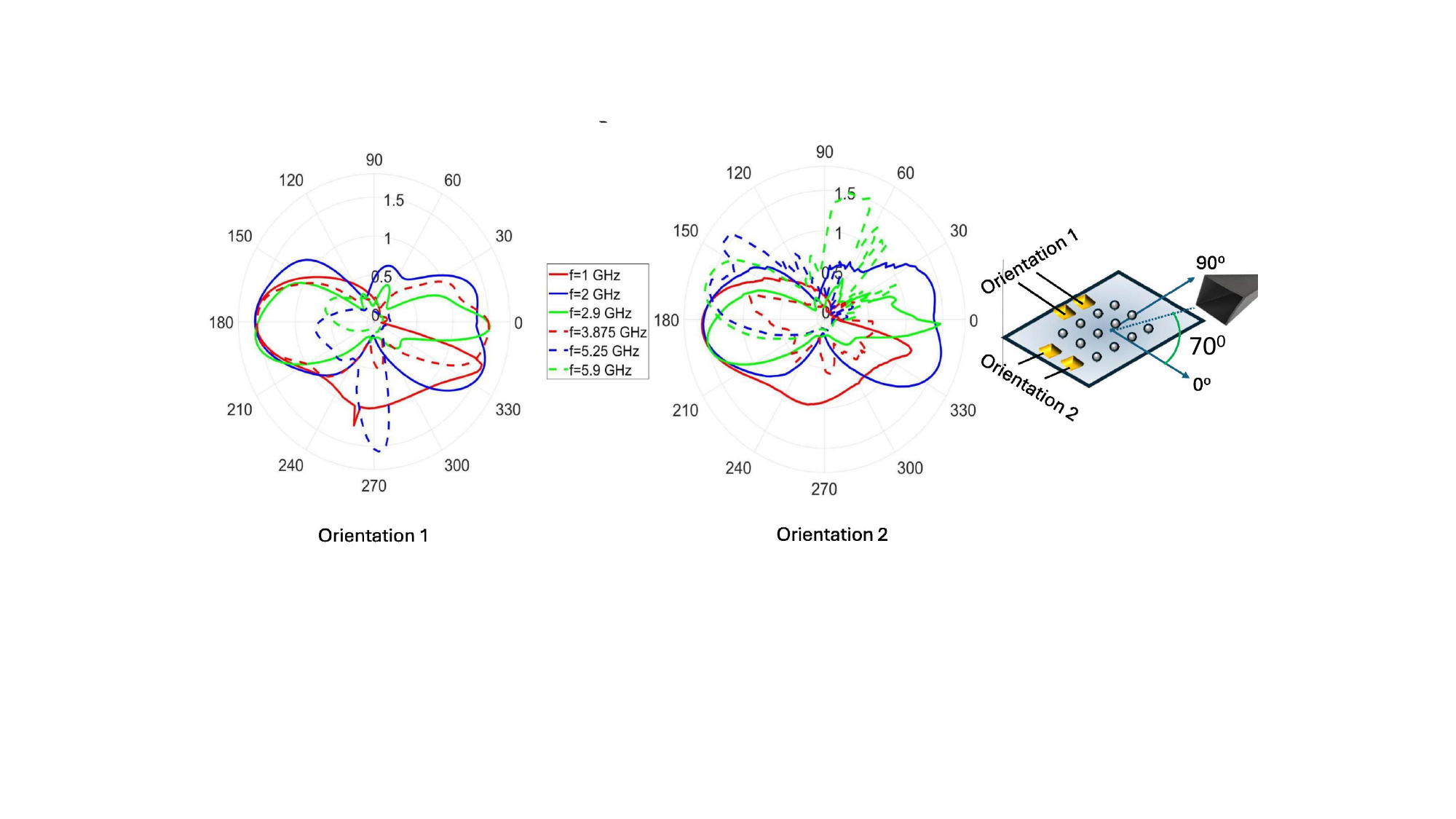}
\caption{Radial plot (in radians) of the polarization angle $\theta$ in the plane of the nanomagnets at different frequencies for two mutually perpendicular directions of SAW propagation labeled as ``orientation 1'' and ``orientation 2''. The far right panel shows that if we place the detector at 70$^{\circ}$ angle in the plane of the nanomagnets, then at 5.9 GHz frequency, we will receive and detect two mutually orthogonal polarizations if we switch the direction of SAW propagation from ``orientation 1'' to ``orientation 2'' by switching the electrode pairs used to launch the SAW.}
\label{fig:Fig4}
\end{figure*}

\subsection{Polarization dependence on the direction of SAW propagation}

In Fig. \ref{fig:Fig3}, we first plot the radiation patterns in the plane of the nanomagnets for both vertical and horizontal polarizations in absolute units (as opposed to in dbi) for both directions of SAW propagation. This is done for different frequencies.

Let us say that at a given frequency the horizontal polarization value in a given direction  is $h$ and the vertical polarization value is $v$. We define a polarization angle $\theta$ as 
\begin{equation*}
    \theta = tan^{-1} \left ( \frac{v}{h} \right ).
\end{equation*}

In this case, $\theta$ = 0 corresponds to horizontal polarization and $\theta$ = $\pi/2$ radians corresponds to vertical polarization.

In Fig. \ref{fig:Fig4}, we show the radial plot of $\theta$ at different frequencies in radians. Note that at 5.9 GHz, in the 70$^{\circ}$ direction, $\theta$ is nearly $\pi/2$ radians for orientation 2 and nearly 0 radians for orientation 1 (see also Fig. s6 in the Supporting Information). Thus, by switching the direction of SAW propagation from orientation 1 to orientation 2, we can change the polarization of the beam radiated in the 70$^{\circ}$ direction at 5.9 GHz (in the nanomagnets' plane) from nearly {\it horizontal} to nearly {\it vertical}.  We can place the receiver at 70$^{\circ}$ angle as shown in the right panel of Fig. \ref{fig:Fig4} and receive either horizontal polarization or vertical polarization by switching the direction of SAW propagation, i.e.,   by switching the electrode pairs that are activated to launch the SAW. This is the basis of the polarization switch for digital communication. We can transmit bits 0 and 1, encoded in horizontal and vertical polarizations, respectively, by switching between two pairs of electrodes. Note that the positions of the transmitting device and the receiver do not have to be fixed, but their relative alignment have to be fixed in this scheme. This enables {\it point-to-point} communication via polarization.

Take another frequency of 5.25 GHz and the 275$^{\circ}$ direction. For orientation 1, $\theta$ is very close to $\pi$/2 (again see Fig. S6 in the Supporting Information), meaning that the emitted beam is vertically polarized whereas for orientation 2, $\theta$ is close to 0, meaning that the emitted beam is horizontally polarized. This is the opposite of the previous case because here orientation 1 corresponds to bit 1 and orientation 2 corresponds to bit 0. Thus, there are multiple directions and frequencies at which point to point communication is enabled via polarization encoding. This feature also lends itself to {\it mutiple-input-multiple-output} (MIMO) antenna functionality. Multiple bit streams, each with its own carrier frequency, can be transmitted in multiple different directions simultaneously to improve reliability, countering signal fading, etc.

The specific directions and the specific frequencies for communication (70$^{\circ}$ and 5.9 GHz or 275$^{\circ}$ and 5.25 GHz in this case) will depend on the antenna parameters such as size, shape, spacing, material composition, etc. of the nanomagnets and can be changed by changing these parameters. However, to develop a theory and simulation framework for designing the nanomagnet array to obtain a specific direction at a specific frequency is a major undertaking in itself and beyond the scope of this work. Fortunately, this is not needed for the applications mentioned here. One will simply measure the radiation pattern at different frequencies and experimentally obtain the correct directions and frequencies for communication. {\it This is a calibration exercise -- every antenna is calibrated}. Once an antenna is calibrated, it is ready for use. 

\subsection{Covert communication}

The fact that the point-to-point communication via polarization encoding works only for {\it a specific alignment of
the transmitter and receiver} makes {\it covert} communication \cite{jiang,nan} possible. The authorized sender has access to the polarization
radial plot and can inform the authorized receiver where to place the receiving device. An unauthorized receiver,
intent on intercepting the message (eavesdropping), will not know where to place the receiver and will place it at a
wrong location with very high probability. The eavesdropper will therefore not receive binary polarization states and
hence cannot eavesdrop or intercept the message. This is similar to {\it steganography} \cite{fridrich,grabsky} and eliminates the need for encryption or cryptography which are
always vulnerable to sophisticated attacks.

\subsection{Physical unclonable function}

The scheme is also unclonable since the specific alignments and frequencies that work will vary slightly from sample to sample because
of unavoidable manufacturing variations involved in the fabrication of the AMMC. Thus, the correct alignment direction and frequency are a ``biometric'' of any chosen antenna and
hence can make the antenna act as a physical unclonable function (PUF) that cannot be reproduced or predicted \cite{devadas}, thereby providing
strong security.

\subsection{Stealthy antenna}

Finally, the above feature introduces an element of {\it stealth} in the antenna operation. Correct bit streams are transmitted only the right direction at the right frequency and an eavesdropper whose receiver is not placed in the right location and tuned to the right frequency will receive garbled bit streams. For example, at 5.9 GHz, $\theta$ = 0 for both directions of SAW propagation (both orientations 1 and 2) in the 340$^{\circ}$ direction. Hence, a receiver placed in that 340$^{\circ}$ direction will receive a constant stream of 0's if the horizontal polarization encodes the bit 0. This bit stream contains no message. Similarly, in the 165$^{\circ}$ direction, $\theta$ = $\pi/2$ radians for orientation 2 and 0.61 radians for orientation 1 at 5.9 GHz. Thus, the beam is 100\% vertically polarized for the choice of orientation 2 while being 67\% horizontally polarized and 33\% vertically polarized for the choice of orientation 1 if the chosen frequency is again 5.9 GHz. The latter impure polarization states do not encode binary bits unambiguously and hence carry no meaningful message.  Thus, the message is concealed from anyone without precise knowledge of the correct orientation and frequency. This enables stealth.

\section{Conclusion}

We have experimentally demonstrated a novel polarization switch antenna for microwave signals (1-6 GHz) that is nearly three orders of magnitude {\it smaller} than the electromagnetic wavelength at the operating frequency. That enables miniaturization of the antenna which will allow it to be integrated on a chip. Such an antenna, which, to our knowledge, is demonstrated here for the first time, can be used as an ultra-compact polarization division multiplexer and to transmit digital data encoded in two mutually orthogonal polarizations at microwave frequencies from point-to-point, while sporting some unique features that can be leveraged for covert and stealthy communication. It can be used to construct a physically unclonable function (PUF) for electromagnetic authentication. These additional attributes enhance the appeal of this implementation.

\section*{Acknowledgment} The authors are indebted to Dr. Jonathan Lundquist for help with antenna measurements.

\section*{Supporting Information}
Supporting information can be found at the Wiley Online Library or from the corresponding author.

\section*{Conflict of Interest}
The authors declare no conflict of interest.

\section*{Data Availability Statement}
The data that support the ﬁndings of this study are available from the corresponding author upon reasonable request.


\begin{thebibliography}{1}
\bibitem{oe}
H. Yuan, B. Zhang, Z. Zhong,
{\it Opt. Express}, {\bf 2023}, {\it 31}, 43034.
\bibitem{li}
Z. Li, D. Zhang, J. Liu, J. Zhang,  L. Shao, X. Wang, R, Jin, W. Zhu, {\it Adv. Photonic Res.}, {\bf 2021}, {\it 2}, 2100175.
\bibitem{staacke}
R. Staacke, R. John, M. Kneiss, C. Osterkamp, S. Diziain, F. Jelezko, M. Grundmann and J. Meijer,  {\it J. Appl. Phys.}, {\bf 2020}, {\it 128}, 194301. This is not an antenna and hence does not have applications in digital long-distance communication.
\bibitem{jin}
J. Zhang, Z. Li, C. Zhang, L. Shao and W. Zhu, {\it npj 2D materials and applications}, {\bf 2022}, {\it 6}, 47. This methodology cannot create two orthogonal polarizations, but can rotate the polarization of an incident microwave beam by up to 45$^{\circ}$. Thus, it is not suitable for binary encoding.
\bibitem{imai}
T Imai, K. Nosu, H. Yamaguchi, {\it Electron. Lett.}, {\bf 1985}, {\it 21}, 52. 
\bibitem{collett}
E. Collet, {\it Field Guide to Polarization}. (SPIE Digital Library, 2005). doi.org/1.0.1117/3.626141.
\bibitem{holmes}
D. Holmes, {\it J. Opt. Soc. Am.}, {\bf 1964},  {\it 54}, 1340. 
\bibitem{lefevre}
H. C. Lefevre,  {\it Electron. Lett.}, {\bf 1980}, {\it 16}, 778. 
\bibitem{okoshi}
T. Okoshi, Y. H. Cheng, K. Kikuchi, {\it Electron. Lett.}, {\bf 1985}, {\it 21}, 787. 
\bibitem{lizana}
A. Lizana et al.  {\it Opt. Lett.}, {\bf 2015}, {\it 40}, 3790. 
\bibitem{kubota}
M. Kubota, T. Oohara, K. Furuya, Y. Suematsu, {\it Electron. Lett.}, {\bf 1980}, {\it 16}, 573. 
\bibitem{zhuang}
Z. Zhuang, S.-W. Suh, J. S. Patel, {\it Opt. Lett.}, {\bf 1999}, {\it 24}, 694. 
\bibitem{she}
A. She, F.  Capasso, {\it Sci. Rep.}, {\bf 2016}, {\it 6}, 26019.
\bibitem{farzin}
P. Farzin,M. J.  Haziahmadi, M. Soleimani, ` {\it Sci. Rep.}, {\bf 2024}, {\it 14}, 11155.
\bibitem{rodrigues}
F. J. Rodríguez-Fortu\~no et al., {\it Laser and Photonics Rev.}, {\bf 2014}, {\it 31}, 27. 
\bibitem{dong}
P. Dong, Y.-K. Chen, G.-H. Duan, D. T. Neilson, {\it Nanophotonics}, {\bf 2014}, {\it 3}, 215.
\bibitem{miller}
D. A. B. Miller, {\it Photonics Res.}, {\bf 2013}, {\it 1}, 1.
\bibitem{brown}
D. E. Browne and T. Rudolph, {\it Phys. Rev. Lett.}, {\bf 2005}, {\it 95}, 010501.
\bibitem{tyo}
J. S. Tyo, D. L. Goldstein, D. Chenault and J. A. Shaw,  {\it Appl. Opt. } {\bf 2006}, {\it 45}, 5453.
\bibitem{raisa}
R. Fabiha, J. D. Lundquist, E. Topsakal, S. Bandyopadhyay, {\it IEEE Trans. Ant. Prop.}, {\bf 2025}, {\it 73}, 2862.

\bibitem{raisa1}
R. Fabiha, J. D. Lundquist, S. Majumder, E. Topsakal, A. Barman, S. Bandyopadhyay, {\it Adv. Sci.}, {\bf 2022}, {\it 9}, 2104644.

\bibitem{nanoscale}
A. De, J. L. Drobitch, S. Majumder, S. Barman, S. Bandyopadhyay, A. Barman, {\it Nanoscale}, {\bf 2021}, {\it 13}, 10016.

\bibitem{saibal}
A. Samanta, S. Roy, {\it IEEE Trans. Electron Devices}, {\bf 2023}, {\it 70}, 
335.
\bibitem{carman}
J. P. Domann, G. P. Carman, {\it  Appl.
Phys.}, {\bf 2017}, {\it 121},  044905.
\bibitem{arxiv1}
R. Fabiha, M. Suche, E. Topsakal, P. J. Taylor, S. Bandyopadhyay, {\bf 2025}, arXiv preprint arXiv:2408.16854.
\bibitem{arxiv2}
R. Fabiha, P. K. Pal, M. Suche, A. K. Mondal, E. Topsakal, A. Barman, S. Bandyopadhyay, {\bf 2025}, arXiv preprint arXiv:2408.08368.

\bibitem{Harrington}
R. F. Harrington,  {\it J. Res. Nat. Bur. Stand.}, 1960, {\it 64}, 1-12.
\bibitem{skrivervik}
A. K. Skrivervik, J. F.  Z\"urcher, O. Staub and J. R. Mosig,  {\it IEEE Antennas
Propag. Mag.}, 2001, {\it 43}, 12.
\bibitem{Chu}
L. J. Chu, {\it J, Appl. Phys.}, 1948, {\it 19}, 1163.
\bibitem{UCLA}
M. N. S. Prasad, Y. Huang and Y. E. Wang, 2017 XXXIInd General Assembly and Scientific Symposium of the International Union of Radio Science (URSI GASS), DOI: 10.23919/URSIGASS.2017.8105322.
\bibitem{jiang}
Y. Jiang, L. Wang, H-H Chen and X. Shen, {\it Proc. IEEE}, 2024, {\it 112}, 47.
\bibitem{nan}
X. Chen, J. An, Z, Xiong, C. Xing, 
N. Zhao, F. R. Yu and A, Nallanathan, {\it IEEE Commun. Surv. Tut.}, 2023, {\it 25}, 1173.
\bibitem{fridrich}
J. Fridrich, {\it Steganography in Digital Media:
Principles, Algorithms, and
Applications}.  Cambridge
Univ. Press, (Cambridge, UK) 2009.
\bibitem{grabsky}
S. Grabski and K. Szczypiorski, {\it 
Proc. IEEE Secur. Privacy Workshops}, 2013, 158.
\bibitem{devadas}
C. Herder, M-D Yu, F. Koushanfar and S. Devadas, {\it Proc. IEEE}, 2014, {\it 102}, 1126.
\end{thebibliography}
\end{document}